\documentstyle[manuscript,aps]{revtex}
\tightenlines

\begin{document}

\title{Stability of gold nanowires at large Au-Au separations} 

\author{N.V. Skorodumova$^1$ and S.I. Simak$^2$}
\address{
$^1$ Department of Materials Chemistry, Uppsala University, Box 538, S-75121, Uppsala, Sweden\\
$^2$ Department of Applied Physics, Chalmers University of Technology and Gothenburg University, 
S-41296, Gothenburg, Sweden} 

\maketitle

\begin{abstract} 
   The unusual structural stability of gold nanowires at large separations of 
gold atoms is explained from first-principles quantum mechanical calculations.
We show that undetected light atoms, in particular  hydrogen, stabilize the 
experimentally observed structures, which would be unstable in pure gold 
wires. The enhanced cohesion is due to the partial charge transfer from gold 
to the light atoms. This finding should resolve a long-standing controversy 
between theoretical predictions and experimental observations.
\end{abstract}
\date{\today}

\newpage
\setcounter {page}{1}

Gold is known as the noblest metal of all due to its inert behavior as a 
bulk material and low reactivity of its perfect surfaces. This property 
made gold probably the most valuable and desirable metal in modern human 
history. At the same time, it still challenges scientists with some puzzling 
features. Among them there are striking structural and electrical properties of 
gold nanocontacts. Such nanocontacts are obtained by the mechanically 
controllable break-junction (MCB) technique or by the scanning tunneling 
microscopy (STM), when the tip is first driven into the gold sample, and 
then retracted forming a nanowire \cite{kondo,ohnishi,yanson}. 
On stretching the nanowire undergoes  
a number of structural rearrangements finally producing a monoatomic gold 
bridge suspended between the tip and sample. In most experiments such nanowires 
are found to be amazingly stable, reaching the interatomic distances of about 3.6 - 4.0 $\AA$ 
\cite{ohnishi,yanson,rodrigues} i.e. significantly larger than the nearest-neighbor distance in bulk 
gold (2.88 $\AA$). 
The conductance of the nanocontacts is quantized in steps of 
$2e^2/h$ approaching the theoretical one-dimensional limit of $2e^2/h$ when the 
suspended wire becomes monoatomic \cite{ohnishi,yanson}. 
A different approach to produce and study gold nanowires is based
on the transmission electron microscopy (TEM). In this technique 
gold nanowires are fabricated by perforating holes in a gold thin film using a
focused electron beam with current density $\sim$ 100 $A/cm^{2}$. 
The structural reorganization of the nanowire is further studied by the electron beam with current
density either reduced to $\sim$ 30 $A/cm^{2}$  \cite{ohnishi,rodrigues} or kept the same as
in the preparation process \cite{takai}.
The results of the former experiments \cite{ohnishi,rodrigues} generally agree with 
the results of the STM and MCB based experiments, finding long gold interatomic 
distances of 3.4 - 3.6 $\AA$ \cite{rodrigues}.
On the contrary, the latter experiment \cite{takai} does not confirm the existence of long Au-Au bonds,
reporting the bulk-like spacing of $\sim$ 2.9 $\AA$.

The issue of the novel gold nanowires has already been addressed in a large body of theoretical 
and experimental works 
\cite{landman,sanchez,maria,skoro,rubio,rodrigues2,kondo99} which describe many important features of 
gold nanocontacts, 
including their formation, structural rearrangements and conducting 
properties. However, a convincing explanation of the puzzling structural 
stability of gold monoatomic wires at large Au - Au separations, seen in most 
experiments, is still lacking. In the present report 
we show from first-principles quantum mechanical calculations that  
the unusual structural stability of gold nanowires at large separations of 
gold atoms can be explained by the presence of undetected light atoms 
able to stabilize the
experimentally observed structures, which would be unstable in pure gold
wires. In particular, we find hydrogen to be a primary suspect. 

To understand the origin of the phenomena we have performed 
first-principles calculations in the framework of the density functional theory (DFT) 
\cite{method}. 
Two sets of data have been calculated at T = 0 K, first, for the gold 
nanowires supported by gold tips consisted of 28 atoms, with monoatomic 
part consisted of 2 to 6 atoms, and second, for infinite monoatomic chains 
with the similar number of gold atoms treated as inequivalent in the 
periodically reproduced unit cells. 
In agreement with 
previous findings \cite{sanchez,rubio}, the results obtained for both sets 
(with and without supporting tips) are very similar, 
indicating that the stability of a nanocontact is essentially determined 
by its monoatomic part. 
No magnetization is found for any studied configuration. 
The binding energy curve for an infinite gold
monoatomic wire is given in Fig. 1(a) together with the equilibrium atomic 
configurations and the ranges of their stability. 
Zigzag is an energetically favorable atomic configuration. 
Upon stretching the wire becomes linear at the Au-Au 
distance of about 2.7 $\AA$ remaining stable only in a narrow interval up to 
about 2.9 $\AA$, when the Peierls distortion triggers the dimerization of the 
wire accompanied by a metal-insulator transition  \cite{skoro} (Fig. 1(a)). 
Decreasing the Au-Au distance below the range shown in Fig. 1(a) leads 
to a series of deepening local energy minima and corresponds to the chain 
folding into the bulk-like structure.

These results suggest the
wire to break (and obviously stop being conductive) already at interatomic 
distances of
2.9 $\AA$,
in agreement with previous calculations 
\cite{sanchez,maria,skoro,rubio} and the experiment by Takai {\it et. al} \cite{takai}. 
They, however, by no means explain the  stability of gold monoatomic wires 
with the spacing of 3.4 - 4.0 $\AA$  observed in most of other experiments 
\cite{ohnishi,yanson,rodrigues}. 
Therefore the existing first-principles theory does not
provide a complete description of the phenomenon. It is especially remarkable regarding the fact 
that very different gold systems (bulk, surface, dimers, atomic clusters), 
and in particular their geometry, are known to be accurately described in 
the framework of the DFT \cite{AU}. It seems to be a reasonable 
question to ask if there could be anything missing in the previous theoretical considerations
 of gold nanowires. For instance,
it is tempting to check
whether small amounts of light atoms, which are 
hard to detect in standard experiments, may influence the stability of gold 
nanocontacts. 

We start with an observation that very similar results for the Au-Au spacing 
(3.4 - 4.0 $\AA$) have been obtained in different experiments, with and without 
(ultra) high vacuum conditions, {\it in situ} sample preparation, and controlled 
low temperature \cite{ohnishi,yanson,rodrigues,ohnishi2}. Especially in the 
case of the ultra-high vacuum the 
probability to encounter most of the light atoms, such as, for instance, carbon and oxygen, 
is expected to be low. 
In particular it has been shown by Ugarte \cite{ugarte} and Rodrigues and Ugarte 
\cite{rodrigues} that in the TEM-based experiments intense electron
irradiation is helpful against carbon contamination. 
Nevertheless, such experiments \cite{rodrigues} still show large Au-Au distances 
in gold nanowires.
There is, however, at least one element, {\it hydrogen}, which is present
even at high vacuum conditions
\cite{abramov,hanlon}. 
It is also known that gold becomes unrecognizably active when 
deposited as small particles on a substrate \cite{valden,bond,sanchez2}. This enhancement of 
gold reactivity is caused by a high concentration of steps and surface 
defects in small particles, as well as by the substrate-induced strain \cite{mavrikakis}. 
Note that such low-dimensional gold structures are well presented on the 
surfaces of the STM tips. Gold surfaces interact with atomic hydrogen, which is 
able to form a stable layer on the surface \cite{cade}. Dissolving into the bulk, 
hydrogen tends to segregate at the surface and subsurface layers \cite{stobinski}. Gold 
has also been reported to play a role of a hydrogen trap in gold alloys that 
may be considered as an indication of a strong interaction between gold and 
hydrogen atoms \cite{pyun}. 

We have extensively studied this interaction from first-principles in the 
systems of different scales: from nanostructures to the fcc bulk gold 
\cite{skor_unpubl}. Our results show that hydrogen 
molecules do dissociate when embedded into the gold matrix. Moreover, being 
dissolved in small quantities in the fcc bulk gold, hydrogen binds strongly 
to Au atoms with characteristic Au-H distances of about 1.75 $\AA$. Moderate 
stress and deformations of the gold host lattice do not break these bonds 
making hydrogen follow Au atoms. 

The most profound influence of the hydrogen presence has been observed in gold 
nanowires. The binding energy curve for a one-dimensional gold-hydrogen wire 
is presented in Fig. 1(b). There is a sequence of structural transformations 
similar to that of a pure gold wire (see Fig. 1(a)). A zigzag arrangement of 
atoms is most stable for Au-Au 
distances smaller than 2.7 $\AA$, followed upon stretching by the linear 
configuration of the gold atoms with zigzag-like positions of the hydrogen 
atoms. This configuration is stable up to about 3.4 $\AA$ of Au-Au distance. 
On further stretching the chain becomes linear with hydrogen placed in 
between the gold atoms reaching the Au-Au spacing of about 3.8 $\AA$ and Au - H 
spacing of 1.85 $\AA$. Further stretching results in dimerization eventually 
followed by the complete break of the wire. Thus monoatomic gold-hydrogen 
wire shows an enhanced stability at large Au-Au distances that well matches 
the experimentally observed spacing of 3.6 - 4.0 $\AA$  \cite{ohnishi,yanson,rodrigues}. 
The cohesion is 
twice as strong as the one of a pure gold chain (see Fig. 1). 
As experiments on gold nanowires are often performed at room temperature,
we have also studied the stability of monoatomic Au-H wires at finite temperatures
by first-principles molecular dynamics (MD) simulations \cite{MD}. 
According to our MD simulations the structures of the nanowires obtained 
at T = 0 K
remain intact at least up to T = 300 K. 

The nature of bonding in gold nanocontacts containing hydrogen can be 
understood by analyzing the charge density as well as the difference between 
the actual charge density and the atomic densities of gold and hydrogen 
constituting the system (Fig. 2). Charge density presented in Fig. 2(a) reveals 
pronounced electron density bridges caused by hydrogen situated in between 
the gold atoms. This resembles 
the hydrogen-type bonding where the charge transfer from 
Au to H, which can be seen in the charge density difference plot (Fig. 2(b)), 
is a key component.  Therefore, an enhanced effective binding between the 
gold atoms via Au-to-H charge transfer mechanism results in the striking 
stability of the gold-hydrogen nanowires. 

Further, in experiments \cite{ohnishi,yanson} the conductance of gold nanocontacts has been 
measured at a low d.c. bias (about 10 mV). We have also studied the 
stability and ability of the gold-hydrogen chains to conduct at a similar 
bias, which is simulated by shifting the chemical potential at one end of 
the wire with respect to the other in the way described in Ref. \cite{todorov}.  
The band structure of such a wire (Fig. 3) clearly shows the existence of one 
conduction channel, in agreement with theory and most experimental findings 
\cite{ohnishi,yanson,rodrigues}. Additional forces appearing on atoms due to the applied bias are of 
about 0.001 eV/$\AA$  that is by no means sufficient for breaking the bonds. 

In general one cannot rule out the possibility of the presence of other light elements.
First, small amount of light atoms can be already present in the sample used to 
fabricate a nanowire. Second,  not all experiments on gold nanowires employ high vacuum conditions.
We have considered carbon and oxygen as possible contaminants (see Ref.\cite{method} for calculational details)
and found that both of them stabilize monoatomic wires up to the Au-Au distances of $\sim$ 4.5 $\AA$. The enhanced
cohesion is due to the bonding mechanism similar to the one operating in Au-H wires. 
The equilibrium Au-Au distances in Au-O and Au-C wires are, however, somewhat larger 
($\sim$ 3.6 $\AA$ in both systems). Thus we notice that
due to the larger equilibrium Au-Au separation in such systems,
they should be in a "compressed" state in the region of experimentally observed Au-Au distances, 
contrary to the Au - H system. Therefore, if in a particular experiment oxygen or carbon atoms are responsible
for the increased stability of gold nanowires, one would at least expect a different dependence of the force on
the tips displacement from what is actually seen in the STM-based experiments  \cite{rubio}. 

Further, we comment on the experimental results by Takai {\it et al.} \cite{takai} reporting the bulk-like
spacing in gold monoatomic wires and, therefore, confirming the theoretical results for pure
gold wires. We suggest that this experiment was most probably free of contamination 
due to a permanent use of intense electron irradiation (current density $\sim$ 100 $A/cm^2$) during the
observations. As a result, gold nanowires were stable at a bulk-like atomic separation of 2.9 $\AA$ 
within a very short time interval. It is in a sharp contrast with other
experiments \cite{ohnishi,rodrigues}, where an electron beam with current density of 100 $A/cm^2$ was only 
used to perforate neighboring holes in gold films while the observations
were performed at essentially lower current density ($\sim$ 
30 $A/cm^2$).  As we already mentioned, these experiments resulted in long-living 
monoatomic wires with large interatomic distances.

To summarize, we have shown that the striking stability of the suspended gold 
nanowires may be attributed to the hydrogen atoms, which mediate 
gold-gold interactions, dramatically increasing binding. 
Hydrogen, being
 always present but rarely controlled in experiments, alters the properties 
of gold nanowires, the crucial components of nanometer-scale devices 
developed for microelectronics and biophysics applications \cite{mbindyo}.  
In addition we note that other light atoms, like carbon and oxygen, are also able to
stabilize gold nanowires. Our calculations show that Au-O and Au-C monoatomic 
wires are stable up to the Au-Au distances of 4.5 $\AA$.
We suggest that further experiments, incorporating hydrogen and other light 
atoms detection, may lead to important implications for the modern atomic-size technologies. 

We thank M. Springborg, K. Hermannson, Y. Andersson, K. Forsgren, G. K{\"a}ll{\'e}n, 
S. Kubatkin and D. Erts for discussions. Support from the Swedish 
Foundation for Strategic Research and Swedish Research Council is gratefully 
acknowledged.

\newpage
\begin{figure}
\caption{
Binding energy curves for infinite monoatomic wires of pure gold 
(a) and gold with hydrogen (b). The shown energies are given per gold atom 
(a) and gold-hydrogen pair (b) as a function of the average Au-Au distances. 
These are calculated as (d$_1$ +...+ d$_n$)/n, with d$_n$ being the actual distances 
between gold atoms in linear chains or the z-projected distances for 
zigzag-like chains {\protect \cite{skoro}}, where n is the number of gold atoms per unit cell. 
For infinite monoatomic wires the total amount of hydrogen atoms (case (b)) is 
equal to the total amount of gold atoms resulting in 2 to 6 hydrogen atoms 
per simulated nanowire (see Fig. 2).
The energy curves for the chains forced to be strictly 
linear (with all d$_n$ equal) are shown in blue. The equilibrium atomic 
configurations are given in the corresponding ranges of their stability. 
The underlined intervals indicate the range of stability of the chains 
with the linear arrangements of gold atoms.
}
\label{fig.1}
\end{figure}

\begin{figure}
\caption{
Charge density distribution (a) and the difference between the 
actual density and atomic charge densities of gold and hydrogen constituting 
the system (b) for the monoatomic part of the gold-hydrogen nanocontact 
with the Au-Au distance of 3.6 $\AA$. Charge density in (a) spans the range 
0 - 1 electron/$\AA^3$ with the spacing of 0.1 electron/$\AA^3$. Charge density 
difference in (b) spans the range -0.1 - 0.2  electron/$\AA^3$ with the spacing 
of 0.03 electron/$\AA^3$. In both (a) and (b), the color background changing 
from blue via green and red to yellow indicates the increase of the density. 
Accordingly, in the charge density difference plot (b) the blue part of the 
spectrum indicates the lack and the red and yellow ones the excess of 
electrons. Positions of Au and H nuclei are marked with corresponding symbols.
}
\label{fig.2}
\end{figure}

\begin{figure}
\caption{
The band structure of the gold-hydrogen nanowire with the Au-Au 
distance of 3.6  $\AA$ (see Fig. 1)  calculated on a sufficiently dense mesh 
of k-points (30 k-points in both, $\Gamma Z$ and $\Gamma X$ directions).  The energy is 
given with respect to the Fermi level. The perfectly flat bands in $\Gamma X$ 
direction indicate the absence of interaction between single nanowires 
simulated in three-dimensional periodic supercells.
}
\label{fig.3}
\end{figure}


\begin{thebibliography}{99}
\bibitem{kondo}
Y. Kondo, K. Takayanagi, Science {\bf 289}, 606 (2000).
\bibitem{ohnishi}
H. Ohnishi, Y. Kondo, K. Takayanagi, Nature {\bf 395}, 780 (1998). 
\bibitem{yanson}
A. I. Yanson,  G. R. Bollinger, H. E. van den Brom, N. Agra{\"{\i}}t, J. M. van Ruitenbeek, 
Nature {\bf 395}, 783 (1998).
\bibitem{rodrigues}
V. Rodrigues and D. Ugarte, Phys. Rev. B {\bf 63}, 073405 (2001).
\bibitem{takai}
Y.Takai, T. Kawasaki, Y. Kimura, T. Ikuta, and R. Shimizu, Phys. Rev. Lett. {\bf 87},
106105-1 (2001).
\bibitem{landman}
U. Landman, W. D. Luedtke, B. E. Salisbury, and R. L. Whetten, Phys. Rev. Lett. {\bf 77}, 1362 (1996).
\bibitem{sanchez}
D. S{\'a}nchez-Portal et al., Phys. Rev. Lett.  {\bf 83}, 3884 (1999).
\bibitem{maria}
L. De Maria, M. Springborg, Chem. Phys. Lett. {\bf 323}, 293 (2000).
\bibitem{skoro}
N. V. Skorodumova and S. I. Simak, Comp. Mat. Sci. {\bf 17}, 178 (2000). 
\bibitem{rubio}
G. Rubio-Bollinger, S. R. Bahn, N. Agra{\"{\i}}t, K. W. Jacobsen, S. Vieira, Phys. Rev. Lett. {\bf 87}, 026101-1 (2001).
\bibitem{rodrigues2}
V. Rodrigues, T. Fuhrer, D. Ugarte, Phys. Rev. Lett.  {\bf 85}, 4124 (2000);
\bibitem{kondo99}
Y. Kondo,
H. Kimata,
H. Ohnishi,
K. Takayanagi, J. Electron Microsc. {\bf 48}, 1081 (1999).
\bibitem{method}
Calculations were done by 
all-electron projector augmented-wave (PAW) method \cite{kresse,blochl} based on the 
DFT within the generalized gradient approximation (GGA) \cite{perdew}. The cut-off 
of 312 eV for Au and Au - H systems, and 500 eV for Au - C and Au - O were used. 
One-dimensional chains were simulated by the three 
dimensional periodic tetragonal supercells with chains along z-axis and 
separated in x and y directions by 14 $\AA$. Hellmann-Feynmann forces were 
systematically calculated, and the nuclei steadily relaxed to equilibrium 
positions. The integration over the Brillouin zone was performed on a 1x1x24 
k-point mesh. According to the convergence tests, this was a relevant 
amount for an accurate description (within 1 meV per atom) of the energy 
differences in question.
\bibitem{AU} In particular, with the method used here (see Ref. \cite{method}) the lattice
parameter of the bulk gold is reproduced within 1 \% accuracy. An example of 
DFT calculation of gold  surfaces can be found in Ref. \cite{norskov}.
Discussion of the properties of gold clusters of different size 
calculated within DFT can be, for instance, found in Refs. \cite{landman,hakkinen,haberlen} 
\bibitem{ohnishi2}
H. Ohnishi, Y. Kondo, K. Takayanagi, Surf. Sci. {\bf 415}, L1061 (1998). 
\bibitem{ugarte}
D. Ugarte, Nature {\bf 359}, 707 (1992);
D. Ugarte, Chem. Phys. Lett. {\bf 209}, 99 (1993).
\bibitem{abramov}
E. Abramov,  D. Eliezer, Hydrogen Effects in Materials. Edited by A. W. Thompson and N. R. Moody, 295 (1996). 
\bibitem{hanlon}
J. F. O'Hanlon, A User's Guide to Vacuum Technology. John Wiley \& Sons,  New York, 387-396 (1989). 
\bibitem{valden}
M. Valden, X. Lai, D. W. Goodman, Science {\bf 281}, 1647 (1998). 
\bibitem{bond}
G. C. Bond, Catal. Rev.-Sci. Eng. {\bf 41(3\&4)}, 319 (1999). 
\bibitem{sanchez2}
A. Sanchez et al., J. Phys. Chem. {\bf 103}, 9573 (1999). 
\bibitem{mavrikakis}
M. Mavrikakis, P. Stoltze and J. K. N{\o}rskov, Catal. Lett. {\bf 64}, 101 (2000). 
\bibitem{cade}
I. {\v C}ade{\v z}, R. I. Hall, M. Landau, F. Pichou, C. Schermann,  J. Chem. Phys.  {\bf 106}, 4745 (1997). 
\bibitem{stobinski}
L. Stobi{\' n}ski and R. Du{\'s}, Appl.  Surf. Sci.  {\bf 62}, 77 (1992). 
\bibitem{pyun}
S. Pyun, W. Lee, T. Yang, Thin Solid Films {\bf 311}, 183 (1997).
\bibitem{skor_unpubl} N. V. Skorodumova and S. I. Simak (to be published).
\bibitem{MD} 
The first-principles molecular dynamics (MD) simulations, as implemented
in Ref. \cite{kresse} were performed
in the canonical ensemble using the Nos{\'e} thermostat \cite{Nose} for
temperature control.
The forces acting on the atoms were calculated from the ground-state
electronic
energies according to the Hellmann-Feynman theorem at each time step and
subsequently
used in the integration of Newton's equation of motion.
The temperature of 300 K was used throughout the MD simulations.
\bibitem{todorov}
T. N. Todorov,  J. Hoekstra, and A. P. Sutton, Phil. Mag. B  {\bf 80}, 421 (2000).
\bibitem{mbindyo}
J. K. N. Mbindyo et al., Adv. Mat. {\bf 13}, 249 (2001).
\bibitem{kresse}
G. Kresse and J. Furthm\"uller, Comp. Mater. Sci. {\bf 6}, 15  (1996); 
G. Kresse and J. Furthm\"uller, Phys. Rev. B  {\bf 54}, 11169 (1996); 
G. Kresse and J. Joubert,  Phys. Rev. B {\bf 59}, 1758  (1999).
\bibitem{blochl}
P.E. Bl\"ochl, Phys. Rev. B  {\bf 50}, 17953 (1994). 
\bibitem{perdew}
J.  P. Perdew et al. Phys. Rev. B {\bf 46},  6671 (1992). 
\bibitem{Nose} S. J. Nos{\'e}, Chem. Phys. {\bf 81}, 511 (1984).
\bibitem{norskov}
B. Hammer and J.K. N{\o}rskov,  Nature {\bf 376}, 238 (1995).
\bibitem{hakkinen}
H. H\"akkinen and  U. Landman, Phys. Rev. B {\bf 62}, R2287 (2000). 
\bibitem{haberlen}
O. D. H\"aberlen, S. Chung, M. Stener, N. R\"osch, J. Chem. Phys {\bf 106}, 5189 (1997). 
\end{thebibliography}
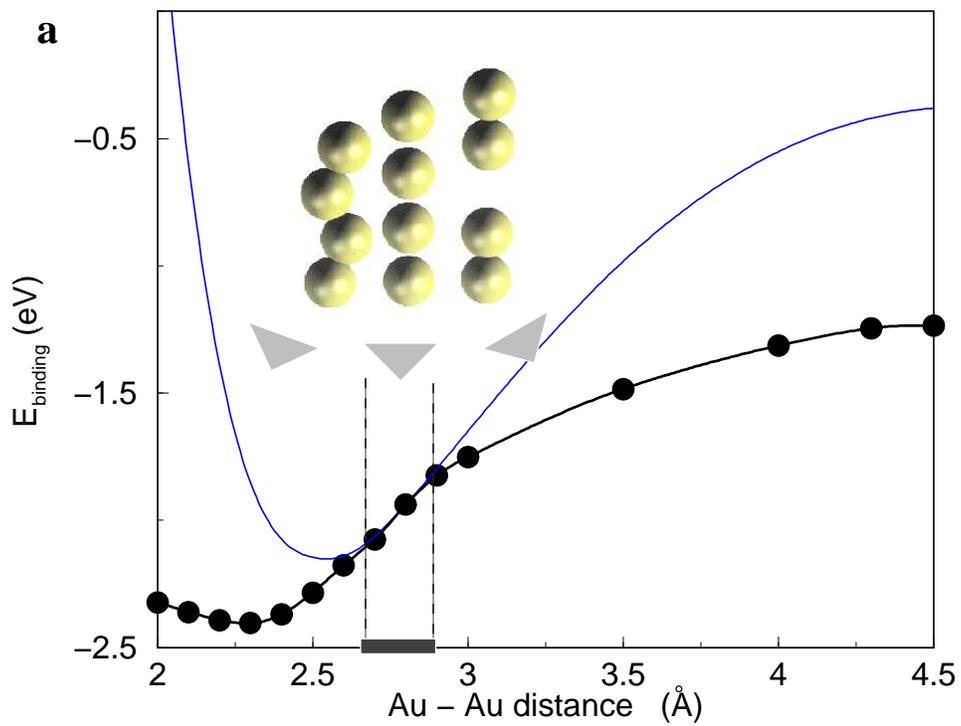
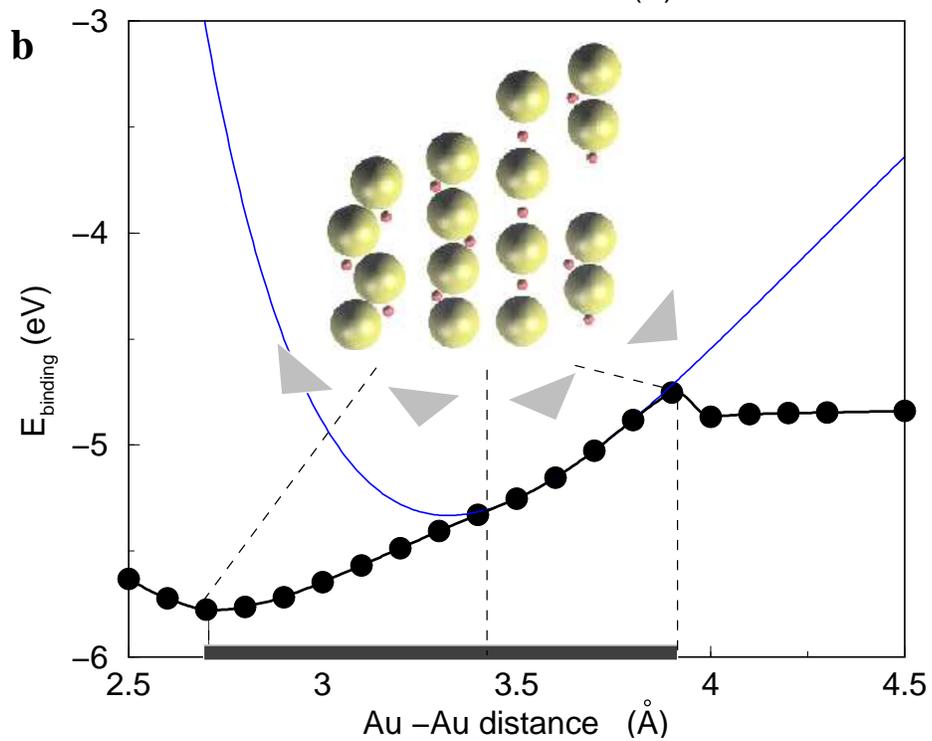
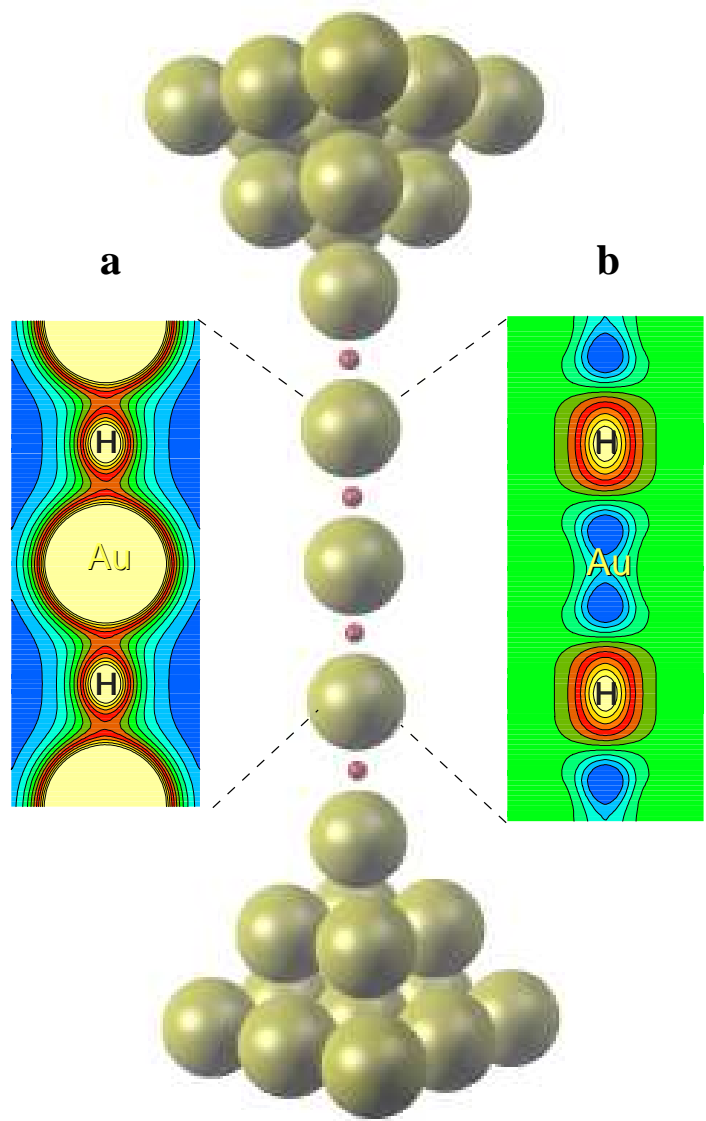
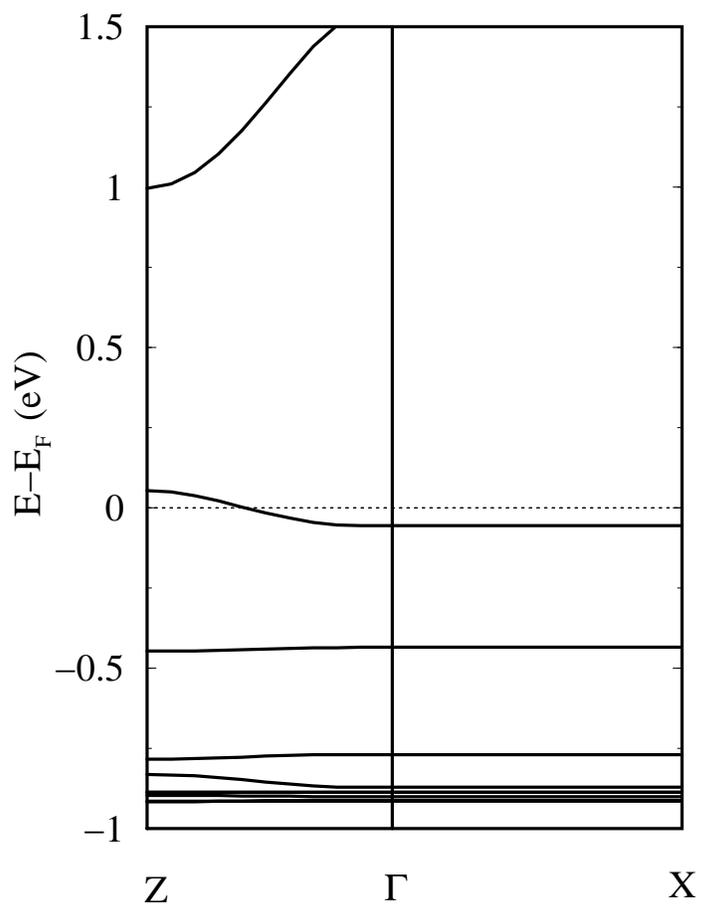
\end{document}